\documentclass{iaus}    
\usepackage{graphicx}    
    
\title[Middle Age SNe] %% give here short title %%    
{Supernovae astrophysics from Middle Age documents}    
    
\author[Polcaro \& Martocchia]   %% give here short author list %%    
{V. F. Polcaro$^1$%    
  \break \and A. Martocchia$^2$ \thanks{Present address: CESR, Toulouse, France},    
 }    
    
\affiliation{$^1$IASF-INAF, Rome, Italy \break email: polcaro@rm.iasf.cnr.it\\[\affilskip]    
$^2$OAS, Strasbourg, France \break email: martok@quasar.u-strasbg.fr }    
    
\pubyear{2005}    
\volume{230}  %% insert here IAU Symposium No.    
\date{?? and in revised form ??}    
\jname{Proceedings Populations of High-Energy Sources in Galaxies}    
\editors{ Evert J A Meurs  \& Giuseppina Fabbiano, eds.}    
\begin{document}    
    
\maketitle    
    
\begin{abstract}    
The supernova explosion of 1054 AD, which originated the Crab Nebula and    
Pulsar, is probably the astronomical event which has been   
most deeply studied by means of   
historical sources. However, many mysteries and    
inconsistencies, both among the different sources and between what is    
deduced by the historical records and the present day astronomical data,    
are demanding extraordinary efforts by theoretical astrophysicists  
in order to put all the data in a meaningful framework.    
An accurate analysis of the 
%%%most significant 
historical sources,   
like the one we are presenting here, may contribute to solve some   
of these problems.    
\end{abstract}    
  
Galactic Supernovae are rare events and their testimonies are extremely important for  
astrophysics.    
To date, seven astronomical events documented by historical texts are   
believed to have been galactic supernovae (see Tab.1).   
Information gathered from the historical    
sources concerning all of these events have been used in some way in astrophysical studies,    
though it is not always easy to extract quantitative data from ancient measurements.    
  
Before the Cepheid distance scale was extended by HST to include the host galaxies of many     
SNe of Type Ia with good photometry, historical Type Ia SNe had been used to compute the     
Hubble constant by combining their brightness at maximum with the modern distances to their     
remnant in order to estimate the peak absolute magnitude. For instance, Schaefer (1996),    
from  a careful reconstruction of the Type Ia SN\,1572 light curve,  obtained a  peak     
magnitude of V=-4.53$\pm$0.18, corresponding to an absolute magnitude of V$_o$=-18.64$\pm$0.31,    
deriving H$_o$=66$\pm$12 km s$^{-1}$ Mpc$^{-1}$: this is an     
astonishing precise result, considering that it was obtained by using data gathered by naked     
eye more than 400 years ago!    
    
However, 
it may be risky to use historical astronomical data for astrophysics if they    
have not been carefully checked from a historical point of view. The most indicative case is     
the one of SN\,185, which was used for calibrating the brightness of     
Type Ia SNe and hence for deriving the Hubble constant. 
%%%However, 
But, as it has been shown by Schaefer (1995), most likely this event was {\it not}   
a supernova, but a transit of comet P/Swift-Tuttle -- indeed the derived value of H$_o$ was     
the unrealistic one of H$_o$$\simeq$ 150 km s$^{-1}$ Mpc$^{-1}$.    
  
  \begin{table}\def~{\hphantom{0}}    
  \begin{center}    
  \caption{Possible historical Supernovae}    
  \label{tab:kd}    
\hskip-8mm  
  \begin{tabular}{lccccl}\hline    
Year    	& Date         & mag   & Remnant            & SN Type           & Source(s) \\  
\hline%\\%[3pt]    
185 AD          & Dec 7(?)     & -2(?) & RWC 86(?)          & Ia(?) Not a SN(?) & Chinese \\    
393 	        & Feb 27-28             & -3(?) & RX J1713.7-3946(?) &                   & Chinese\\    
1006            & Apr 30       & -7.5  & SNR 1006           & Ia   
                                                            & Arabic, Chinese, Japanese, European\\    
1054            & Apr 11 & -4(?) & M1 (Crab)          & II(?) Ib(?)   
                                                            & Chinese, North American(?),\\     
~ &               &              &       &                    & Arabic, Japanese, European \\    
1181            & Aug 6        & -1(?) & 3C 58(?)           & II(?) Ib(?)   
                                                            & Chinese, Japanese, European\\    
1572            & Nov 6        & -4    & Tycho SNR          & Ia   & Tycho Brahe, etc.\\    
1604            & Oct 9        & -3    & Kepler SNR         & Ib(?) & Johannes Kepler, etc.\\  
\hline    
  \end{tabular}    
 \end{center}    
\end{table}    
    
The importance of a careful historical analysis of the historical sources before using     
them for astrophysical purposes is further illustrated by the case of SN\,1054.    
  
  \begin{table}\def~{\hphantom{0}}    
  \begin{center}    
  \caption{The historical records of SN1054 (revised version of Tab.1 from Collins et al., 1999).}    
  \label{tab:1054}    
\hskip-8mm 
\begin{tabular}{lcccccl}\hline    
Date       & MJD  & Ref. & location & appearance as...          & likely $m_{\rm{V}}$ & Notes \\  
\hline%\\%[3pt]    
04/11/1054 & 6126 & a.   & Fustat   & {\it star}                & ~                 & Ibn Butlan \\  
04/11/1054 & 6126 & b.   & Flanders & {\it bright disk at noon} & $\sim -7$         & ~ \\  
04/24/1054 & 6139 & c.   & Ireland  & {\it fiery pillar}        & ~                 & ~ \\  
late April 1054 &~& d.   & Rome     & {\it bright light}        & $< -3.5$          & ~ \\  
05/10/1054 & 6155 & e.   & Liao Kingdom    & {\it star}                & ~ & Sun eclipse \\  
05/14/1054 & 6159 & f.   & Armenia  & {\it star}                & ~                 & ~ \\  
late May 1054(?) &  & g.   & Italy    & {\it very bright star}    & ~                 & ~ \\  
late May 1054   &~& h.   & Japan    & {\it new star... as Jupiter} & $\sim -4.5$    & ~ \\  
June 1054  & ~    & h.   & Japan    & {\it star}                & ~                 & ~ \\  
$\sim$ 06/20/1054 &~&    &     & Crab in conjunction with Sun   & ~ & not visible \\  
07/04/1054 ? & 6210 ? &i.& Song Empire    & {\it star... like Venus}  & $\sim -3.5$       & for 23 days ($^*$) \\  
08/27/1054 ? & 6264 ? &i.& Song Empire    & {\it star... like Venus}  & $\sim -3.5$       & for 23 days ($^*$) \\  
         ~ &    ~ &    ~ &        ~ &                         ~ &                 ~ & ~ \\  
1055       &    ~ & a.   & Constantinople &  {\it star}         &                 ~ & ~ \\  
         ~ &    ~ &    ~ &        ~ &                         ~ &                 ~ & ~ \\  
04/17/1056 & 6863 & i.   & Song Empire    & {\it no more visible}     & $> +6$          & ~ \\  
\hline    
  \end{tabular}    
 \end{center}    
{\tiny JD=MJD+2100000  
  
{\it References:} a. {\it Diary of Ibn Butlan} b. {\it Tractatus de ecclesia} c. {\it Irish Annals}   
d. {\it De Obitu Sancti Leonis} e. {\it Sung-shih hsin-pien} f. {\it Etum Patmich}   
g. {\it Rampona Chronicles} h. {\it Mei Getsuki} i. {\it Sung hui-yao}.   
  
($^*$) Datation may have been falsified (or may just refer to the  
communication to the Emperor), thus to be possibly shifted before. }  
\end{table}    
    
The Song Empire sources were the first to be suggested as witnesses of the birth of the Crab     
Nebula by Hubble (1928) and Mayall (1939).   
It was found that they report a date when the Emperor was notified by the astronomer Yang Weide
about the 1054 "guest star" appearance (4$^{th}$ July), the length of      
the period in which this star was visible in daylight (23 days), the date when the Emperor     
was noticed of the last sighting (17$^{th}$ April 1056), and the star position in  the  sky.     
These data made it possible to prove, though with some problems, the link between 
the historical event and
the explosion of  the precursor of the  Crab  Nebula (Mayall \& Oort, 1942; Duyvendak, 1942).     
Nevertheless, another surely independent Chinese     
source, the {\it K'i-tan-kuo-chih}, the history of the Kingdom of Liao written about 1350 AD,     
refers only briefly to the "guest star", 
stating that the event occurred    
near the time of (or ``during'', following a more recent translation: Collins et al., 1999) a     
total eclipse of the Sun. Duyvendak (1942) has shown    
that  the  only important eclipse of that period occurred on 10$^{th}$ May 1054. This   
evidence of an earlier    
date of the SN\,1054 explosion has been often neglected
because of the contradiction with respect to the date reported by the   
sources of the more developed Song Empire.    
On the other hand, the latter texts give us just two photometric points: on 4$^{th}$ July 1054,    
the star is defined "as luminous as Venus" and thus with V=-3.5$\pm$1; on 17$^{th}$ April     
1056, it is declared "no more visible" and thus having V$\geq$5.5. However, as 
shown by Collins et al. (1999), these two photometric points are not compatible with the    
average light curve of any kind of core-collapse supernova to have exploded on 4$^{th}$   
July 1054: they could be only marginally fitted by the average light curve of a Type Ia   
SN or by a Type II-L SN with its maximum on 20$^{th}$ June.  
While the first hypothesis is ruled out by the presence of the Crab Pulsar,  both  
hypotheses are actually in contradiction with the Song sources themselves,  
stating the appearance of the star on 4$^{th}$ July.   
A problem is that Type II SNe may have significantly different luminosity at peak  
(although their light curves have a very homogeneous behavior when the radioactive 
nuclides decay becomes the dominant energy source: see e.g. Cappellaro \& Turatto, 2001,
for a general review of SN types); however, one cannot but assume that there is  
something  
wrong in the official Song Court report. And, indeed,  
we have nowadays the possibility to consider the whole set of data concerning    
the SN\,1054 event. It is witnessed, to our knowledge,   
by at least 13 primary historical sources all over the world  (see Table 2).     
They are of course much different from 
each other, both in clarity and in style; however, from  
the appearance of the reported phenomena, it is often possible to guess the corresponding  
visual magnitude of the SN\,1054. 
We may therefore try reconstructing a rough light curve, by using the formulas which give   
the naked eye visibility of astronomical objects (see e.g. Schaefer, 1993).    
If we overlap to this curve the observed magnitudes of a few modern Type II SNe, scaled to the    
distance of Crab Nebula (Fig.1, left), and a simple model of Type IIp SN 
with production of 0.07   
M$_{\odot}$ of  $^{56}$Ni (Collins et al, 1999; Sollerman et al., 2001; see Fig.1, 
right), we can notice that the fit is    
fairly good in both cases, though the historical data crowd 
%%%obviously 
in the high-luminosity segment of the curve, as obvious,
making an actual quantitative "best fit" procedure impossible.     
    
We notice that a single point is out from the fit: that is the one related to the very first    
appearance of the SN, obtained from the {\it Tractatus de Ecclesia     
S. Petri Aldenburgensi}, the chronicle of the Church of St. Peter in Oudembourg (in present day Belgium)    
written by an anonymous monk or clerk some 20 years after the reported events. This is the    
text translation, in full (for the latin original see: Guidoboni, Marmo \& Polcaro, 1994):    

\begin{figure}    
\vspace{-4cm}  
\hspace{-2.5cm}  
 \includegraphics{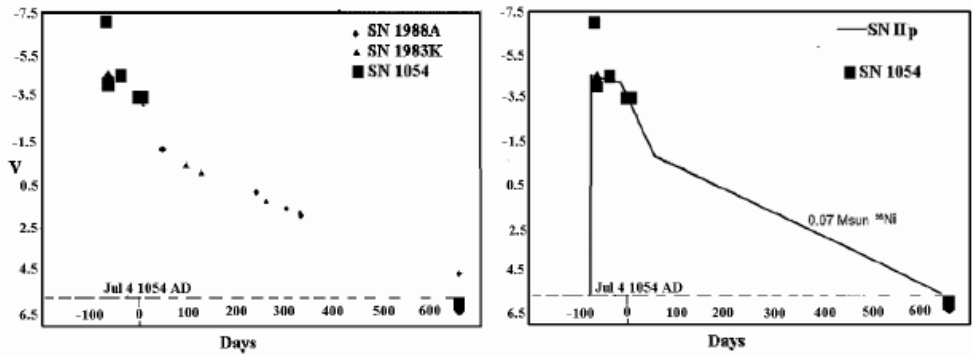}    
  \caption{\small {\it Left:} The historical light curve of the SN\,1054 overlapped to the photometric points  of two modern type II SNe, reduced to the distance of the Crab Nebula. {\it Right:} the
same light curve compared to
a simple model of SN-IIp, assuming a production of 0.07 M$_{\odot}$ of  $^{56}$Ni.}  
\label{fig:lightcurve}    
\end{figure}    
          
{\it And the most  blessed  Pope  Leo,  after  the  beginning  of  the construction   
of the aforementioned church of St.  Peter,  in  the following year, on the  18th day   
before the first of May,  a  Monday,  around midday, happily departed this world.   
And at the same hour as his leaving of the flesh, not only in Rome, where his   
body lies,  but also all over the world there appeared to men a circle in the sky   
of extraordinary brightness which lasted for about half an  hour. Perhaps the Lord   
wished to say that he} [the Pope]   
{\it was  worthy  to receive a crown in Heaven between those who love Him.}    
    
A deep textual analysis of this reference, one of the very few written by an actual eye-witness    
of the SN\,1054, allowed the reconstruction of the exact date of the    
reported phenomenon as the 11$^{th}$ April 1054, in agreement with the date that can be deduced from the    
reports of the Arabic scholar Ibn Butlan (Guidoboni et al, 1994).    
Furthermore, we can observe that    
in this document, the author describes the phenomenon in  neutral     
terms, unaffected by any set of beliefs: the disk-like shape,  the     
intense brightness and the duration of  the  phenomenon  are  all     
elements common to very different cultures. The author  separates     
the description of the phenomenon from  his  cautious      
symbolic  interpretation, 
showing  a  clear  awareness  of   the     
different levels of discourse. The Flemish chronicler saw a bright   
point source at $\sim$30$^o$ from the horizon:    
in the foggy sky of Flanders in Spring, this would appear exactly like a disk.  
Such a short optical transient, in the very first 30 minutes following the collapse    
of a SN-II precursor, could very easily escape the detection even with the   
nowadays observatories;  
therefore, if confirmed by further evidences, this testimony could be very important    
for the understanding of the physics of the core-collapse supernovae.    
      
Thus, if the SN\,1054 exploded on 11$^{th}$ April 1054  
(as Guidoboni, Marmo \& Polcaro, 1992,    
first suggested and Collins et al., 1999, demonstrated) 
it was surely still visible, and very     
impressive, near to the zenith of the Song capital Kaifeng    
during the Sun eclipse of 10$^{th}$ May 1054.   
From this sky configuration and    
by using the standards of the Chinese astrology, which are perfectly documented,  
it is easy to reconstruct the {\it omen} that must have been deduced at the time:    
the Sun represents the Emperor (actually, the Emperor {\it was} the Sun);     
the eclipse is a danger for the Emperor life;  
moreover, the contemporary presence of the "guest star" indicates the loss of the   
Heaven support; therefore the danger is unavoidable: {\it the Emperor will die}.    
It is not surprising that such a terrible omen must have been subject to some form of   
censorship: what the astronomer Yang Weide did -- most probably following the    
wish of Emperor Renzong himself (Polcaro, 2005) -- was to censor all the references to the "guest    
star" preceeding its conjunction with the Sun, which occurred   
at the end of June .  

On the other hand, the Chief Astronomer of the Liao kingdom, whose king actually died one year   
or so after the eclipse, had no reasons to mantain the secret concerning the "guest star"  
visibility during the eclipse and the related omen.  

This is a clear example of the need of taking the historical and   
cultural context into account, to derive meaningful scientific results  
from the study of ancient reports.


\vskip-6mm    
\begin{thebibliography}{00}  
\bibitem[Cappellaro \& Turatto (2001)]{C01}  
  Cappellaro, E., \& Turatto, M. 2001, in: Proc. of the Meeting   
  {\it The influence of binaries on stellar population studies}, Dordrecht:   
  Kluwer Academic Publishers, 2001, Astrophysics and space science library (ASSL),   
  264, 199  
\bibitem[Collins et al. (1999)]{C99}  
 Collins, G.W., Claspy, W.P., Martin, J.C. 1999,   
 \textit{PASP}, 111, 871  
\bibitem[Duyvendak (1942)]{D42}  
 Duyvendak, J.J.L. 1942,   
 \textit{PASP}, 4541, 645.  
\bibitem[Guidoboni, Marmo \& Polcaro (1992)]{G92}  
 Guidoboni, E., Marmo, C., Polcaro, V. F. 1992,   
 \textit{Le Scienze}, 292, 24   
\bibitem[Guidoboni, Marmo \& Polcaro (1994)]{G94}  
 Guidoboni, E., Marmo, C., Polcaro, V. F. 1994,  
 \textit{Mem. SAIt}, 65, 623  
\bibitem[Hubble (1928)]{H28}  
 Hubble, E. 1928,   
 \textit{PASP Leaflet}, 1, 14   
\bibitem[Mayall (1939)]{Ma39}  
 Mayall, N.U. 1939,   
\textit{PASP Leaflet} 3, 119, 145   
\bibitem[Mayall \& Oort (1942)]{MO42}  
 Mayall, N.U., Oort, J.M. 1942,   
 \textit{PASP}, 54, 95   
\bibitem[Polcaro 2005]{P05}  
Polcaro, V.F. 2005, in: \textit{Proc. of SEAC 2005}, Isili, Jun 28--Jul 3, 2005  
\bibitem[Schaefer (1993)]{Sh93}  
  Schaefer, B.E. 1993,  
  \textit{Vistas in Astronomy}, 36, 311  
\bibitem[Schaefer (1995)]{Sh95}  
  Schaefer, B.E. 1995,  
  \textit{AJ}, 110, 1973  
\bibitem[Schaefer (1996)]{Sh96}  
  Schaefer, B.E. 1996,  
  \textit{ApJ}, 459, 438  
\bibitem[Sollerman et al. (2001)]{S01}  
 Sollerman, J., Kozma, C., Lundqvist, P. 2001, \textit{A\&A}, 366, 1971  
\end{thebibliography}
\end{document}